\documentclass[a4paper,11pt]{article}
\pdfoutput=1
\usepackage{jheppub} 
\usepackage[T1]{fontenc} 
\usepackage[english]{babel}
\usepackage[latin1]{inputenc}
\usepackage{amsmath,mathtools}
\usepackage{amsfonts}
\usepackage{amssymb}
\usepackage{wrapfig}
\usepackage{color}
\usepackage{float}
\usepackage{url}
\usepackage{mciteplus}
\usepackage{ifpdf}
\usepackage{epsfig}
\usepackage{xcolor}
\usepackage{hyperref}
\usepackage{cancel}
\usepackage{arydshln}
\usepackage{etoolbox}

\usepackage{xcolor}

\allowdisplaybreaks

\DeclareMathOperator{\Rgod}{\mathnormal{R}_\mathnormal{g}\o\mathnormal{d}}

\renewcommand{\a}{\alpha}
\renewcommand{\b}{\beta}
\newcommand{\g}{\gamma}
\newcommand{\G}{\Gamma}
\renewcommand{\k}{\kappa}

\renewcommand{\o}{\circ}
\newcommand{\Adg}{\text{Ad}_g}
\newcommand{\Adgi}{\text{Ad}_g^{-1}}

\newcommand{\Str}[1]{\text{Str}\left(#1\right)}

\newcommand{\Rg}[1]{\mathnormal{R}_\mathnormal{g}\left(#1\right)}

\newcommand{\dds}{\mathnormal{d}^2\sigma\,}

\newcommand{\diag}{\text{diag}}

\newcommand{\algg}{\mathfrak{g}}
\newcommand{\alggb}{\mathfrak{g}_{\text{b}}}
\newcommand{\alggf}{\mathfrak{g}_{\text{f}}}

\newcommand{\so}{\mathfrak{so}}
\newcommand{\su}{\mathfrak{su}}
\renewcommand{\sl}{\mathfrak{sl}}
\renewcommand{\u}{\mathfrak{u}}
\newcommand{\uosp}{\mathfrak{uosp}}
\newcommand{\psu}{\mathfrak{psu}}
\newcommand{\algi}[1]{\mathfrak{g}^{(#1)}}
\newcommand{\Ai}[1]{A^{(#1)}}

\newcommand{\Z}[1]{\mathbb{Z}_{#1}}
\newcommand{\C}[1]{\mathbb{C}^{#1}}

\newcommand{\comm}[2]{\left[#1,#2\right]}

\newcommand{\Em}[1]{E^{#1}}
\newcommand{\Km}[1]{K_{#1}}

\newcommand{\Cmn}[2]{C_{#1}^{\;\;#2}}

\newcommand{\Lmn}[2]{\Lambda_{#1}^{\;\;#2}}
\newcommand{\dmn}[1]{\delta^{#1}}
\newcommand{\dij}[1]{\delta_{#1}}
\newcommand{\Pb}[1]{P_2\left(#1\right)}
\newcommand{\jm}[1]{j^{#1}}

\newcommand{\ys}[1]{y_{#1}^2}

\newcommand{\dys}[1]{dy_{#1}^2}

\newcommand{\AdS}[1]{AdS_{#1}}
\newcommand{\Sph}[1]{S^{#1}}
\newcommand{\CP}[1]{\mathbb{CP}^{#1}}
\newcommand{\Tpq}[1]{T^{#1}}

\newcommand{\Vnk}[1]{V_{#1}}
\newcommand{\Wnk}[1]{W_{#1}}

\newcommand{\vphi}{\varphi}

\newcommand{\EMa}[1]{\textbf{M}_{#1}}
\newcommand{\EK}[1]{\textbf{K}_{#1}}
\newcommand{\EHi}[1]{\textbf{H}_{#1}}
\newcommand{\EH}{\textbf{H}}
\newcommand{\EM}{\textbf{M}}
\newcommand{\EN}{\textbf{N}}
\newcommand{\EW}{\textbf{W}}
\newcommand{\ET}[1]{\textbf{T}_{#1}}

\newcommand{\EF}[2]{\textbf{F}_{#1}^{(#2)}}
\newcommand{\EL}[2]{\textbf{L}_{#1}^{(#2)}}
\newcommand{\ELpm}[1]{\textbf{L}_{\pm}^{(#1)}}

\newcommand{\es}[1]{\sigma_{#1}}

\renewcommand{\th}[1]{\theta_{#1}}
\newcommand{\dth}[1]{d\theta_{#1}}
\newcommand{\dts}[1]{d\theta^2_{#1}}
\newcommand{\vf}[1]{\varphi_{#1}}
\newcommand{\ph}[1]{\phi_{#1}}
\newcommand{\dvf}[1]{d\varphi_{#1}}
\newcommand{\dps}{d\psi}

\newcommand{\df}[1]{d\phi_{#1}}
\newcommand{\dch}{d\chi}

\newcommand{\dfs}[1]{d\phi_{#1}^2}
\newcommand{\dvfs}[1]{d\varphi_{#1}^2}

\newcommand{\eabc}[1]{\epsilon_{#1}}

\newcommand{\RR}{\mathbb{R}}

\newcommand{\Fd}[1]{\mathcal{F}_{#1}}
\newcommand{\Fdi}[1]{\mathcal{F}^{-1}_{#1}}
\renewcommand{\G}[2]{G\!\left(#1\,|\,#2\right)}
\newcommand{\qs}[1]{q^2_{#1}}

\title{\boldmath 
{Bosonic $\eta$-deformations of Non-integrable Backgrounds}}

\author{Laura Rado,}
\author{Victor O. Rivelles}
\author{and Renato Sánchez}

\affiliation{Instituto de F\'{\i}sica, Universidade de S\~{a}o Paulo \\ Rua do Mat\~{a}o Travessa 1371, 05508-090 S\~{a}o Paulo, SP. Brazil}

\emailAdd{laura@if.usp.br}
\emailAdd{rivelles@fma.if.usp.br}
\emailAdd{renato@if.usp.br}

\abstract{We consider the non-integrable bosonic backgrounds $\Wnk{2,4}\times\Tpq{1,1}$ and $\AdS{5}\times\Tpq{1,1}$ and derive their bosonic $\eta$-deformed versions using an $r$-matrix that solves the modified Yang-Baxter equation obtaining new integrable deformed backgrounds.}

\begin{document}

\makeatletter
\patchcmd{\maketitle}{\@fpheader}{}{\hfill}{} 
\makeatother

\maketitle
\flushbottom

\section{Introduction}
The dynamics of
a field-theoretical system takes the form of differential equations so it is natural 
to ask under which conditions they can be completely solved. Formally, if this happens, 
the system is called integrable. In its Hamiltonian description integrability is ensured by showing the existence of an infinite number of conserved charges which are in involution or, equivalently, by writing its Lax connection. 
The required involution of these conserved charges leads to a particular form of the Poisson bracket of the Lax connection 
in terms of a $r$-matrix \cite{Maillet1985,Maillet1986,Maillet1986a}.
It turns out that this construction is particularly useful when discussing integrability in string theory which was originally applied to the
$\AdS{5}\times\Sph{5}$ superstring  treated as a sigma model on the supercoset $\frac{\psu(2,2|4)}{\so(1,4)\oplus\so(5)}$ \cite{Bena2004,Magro2009}. 

Since integrability is not an ordinary property the exploration of the whole plethora of integrable models is a daunting task. As a consequence, techniques to deform integrable theories while keeping its integrability have been developed. 
The first evidence of integrable deformations was found for the  $SU(2)$ principal chiral model \cite{Cherednik1981} and 
was then generalized for any compact Lie group 
\cite{Klimcik2009}
being now known as Yang-Baxter (YB) sigma models. 
This construction, which is based on a solution of the modified classical Yang-Baxter equation (mCYBE), can be extended to symmetric coset spaces 
\cite{Delduc2013} and to  semi-symmetric coset spaces, as is the case of superstrings in $\AdS{5}\times\Sph{5}$ \cite{Delduc2014,Delduc2014a}
leading to the so called $\eta$-deformed $\AdS{5}\times\Sph{5}$ background \cite{Arutyunov2014,Arutyunov2015}. 
Recently, this kind of deformation was applied to 
the sigma model on the supercoset $\frac{\uosp(2,2|6)}{\so(1,3)\oplus\u(3)}$ which partially describes superstrings in $\AdS{4}\times\CP{3}$ 
\cite{Rado2021}.
Besides that, some instances of integrable deformations of the $\AdS{5}\times\Sph{5}$ and $\AdS{4}\times\CP{3}$ backgrounds were also given based on solutions of the classical Yang-Baxter equation (CYBE) in  \cite{Matsumoto2014,Matsumoto2014a,Matsumoto2015a,Negron2018,Rado2020}. 

More recently it was explored the possibility of starting with non integrable backgrounds and then finding a YB deformation in such a way that the new deformed backgrounds are integrable. That would  give us a powerfull tool to generate  integrable models from non-integrable ones. 
First results in this direction were given for the $\AdS{5}\times\Tpq{1,1}$ \cite{Crichigno2014,Rado2021a} and $\Wnk{2,4}\times\Tpq{1,1}$ \cite{Sakamoto2017} cases. 
In particular, it was found that $\Wnk{2,4}\times\Tpq{1,1}$ gave rise to a somewhat exotic background that is part of a NSNS string background containing a nonvanishing $B$-field \cite{BrittoPacumio1999,PandoZayas2000}. More recently,  $\Tpq{1,1}$ type spaces were studied in the context of new integrable Gaudin models, which can also be used to construct the $\Wnk{2,4}$ spaces, leading to a nonvanishing $B$-field \cite{Arutyunov2020}. However, integrability is lost since $\Wnk{2,4}$ can be realized as the double Wick rotation of $\Tpq{1,1}$ which exhibits a chaotic behaviour \cite{Basu2011}. In this paper we will consider the $\Wnk{2,4}\times\Tpq{1,1}$ and $\AdS{5}\times\Tpq{1,1}$ backgrounds in a first attempt to construct $\eta$-deformations of non integrable spaces in their bosonic sector.

The $\eta$-deformed $\sigma$-model on $\algg$ is given by the action \cite{Delduc2014}  \footnote{We have taken a factor of 1/6 in front of the action, instead of the usual 1/4, in order to obtain the usual undeformed geometry.}
\begin{equation}
\label{Sdef}
S=-\frac{\left(1+c\eta^2\right)^2}{6\left(1-c\eta^2\right)}\int \dds\left(\g^{\a\b}-\kappa\varepsilon^{\a\b}\right)\Str{A_\a,d\,J_\b},
\end{equation}
where $A=-g^{-1}dg\;\in\algg$, $g\in G$, $\g^{\a\b}$ is the string worldsheet metric with $\det\g=1$, and $\k^2=1$. The $\Z{4}$-grading of $\algg$ allows us to split $A$ as
\begin{equation}
A=\Ai{0}+\Ai{1}+\Ai{2}+\Ai{3},\quad \left[\Ai{k},\Ai{m}\right]\subseteq\Ai{k+m}\:\text{mod}\:\Z{4}.
\end{equation}
The operator $d$ is defined as the following composition of projectors $P_i$ $(i=1,2,3)$ on the subalgebras  of $\algg{i}$
\begin{equation}
\label{d1}
d=P_1+\frac{2}{1-c\eta^2}P_2-P_3,
\end{equation}
The absence of $P_0$ is required in order to \eqref{Sdef} be $\algi{0}$-invariant. The deformed current is
\begin{equation}
\label{J1}
J=\frac{1}{1-\eta\Rgod}A,
\end{equation}
where the operator $R_g$ is given by 
\begin{equation}
\label{Rgdef}
\Rg{M}=\Adgi\o R\o\Adg\left(M\right)=g^{-1}R(gMg^{-1})g,\qquad g\in G.
\end{equation}
The operator $R$ is related to the $r$-matrix required for integrability so that the Yang-Baxter equation (YBE) can be written as
\begin{equation}\label{YBE}
\left[RM,RN\right]-R\left(\left[RM,N\right]+\left[M,RN\right]\right)=c\left[M,N\right],\left\{\begin{matrix}
c=0 & \text{CYBE}\\ 
c=\pm 1& \text{mCYBE}
\end{matrix}\right.
\end{equation}
where $M,N\in\algg$. In \eqref{Sdef} and \eqref{YBE} the parameter $c$ refers to either the CYBE equation or to the mCYBE. The $c=1$ case, known as the non-split case \cite{Vicedo2015}, has the $R$ operator \cite{Arutyunov2014}
\begin{equation}
\label{RMij}
R\left(M\right)_{ij}=-i\epsilon_{ij}M_{ij},\qquad \epsilon _{ij}=\left\{\begin{matrix}
1 & \mathrm{if} & i< j\\ 
 0&  \mathrm{if}& i=j\\ 
 -1& \mathrm{if}&i> j 
\end{matrix}\right.,\quad M\in\algg.
\end{equation}

This paper is organized as follows. In \autoref{cosetconstruction} we review the $\Wnk{2,4}\times\Tpq{1,1}$ space and its coset construction as in \cite{Sakamoto2017}. In \autoref{news} we reduce the theory to its bosonic degrees of freedom and apply the $\eta$-deformation discussed above to the bosonic sector of the $\Wnk{2,4}\times\Tpq{1,1}$ background. In \autoref{cosetconstruction2} and \autoref{news2} we review the coset construction and apply the $\eta$-deformation to the bosonic sector of $\AdS{5}\times\Tpq{1,1}$. In \autoref{conclusions} we conclude and discuss our results.

\section{Coset Construction of the Bosonic \texorpdfstring{$\Wnk{2,4}\times\Tpq{1,1}$}{W24xT11} Background \label{cosetconstruction}}

The $\Wnk{2,4}$ space is the noncompact version of the Stiefel space $\Vnk{2,4}$ \cite{James1976}. It is defined by the coset $SO(2,2)/SO(2)$ or equivalently $\left(SL(2,\RR)\times SL(2,\RR)\right)/SO(2)$. 
The $\Tpq{1,1}$ space can be understood as the intersection of a cone and a sphere in $\C{4}$ such that its topology is $\Sph{2}\times\Sph{3}$ and its metric is a $U(1)$ bundle over $\Sph{2}\times\Sph{2}$.  

The isometry group of $\Wnk{2,4}\times\Tpq{1,1}$ is the coset \cite{Sakamoto2017}
\begin{equation}
\label{isometry}
\Wnk{2,4}\times\Tpq{1,1}\equiv\frac{SL(2,\RR)_a\times SL(2,\RR)_b\times SO(2)_r}{U(1)_a\times U(1)_b}\times\frac{SU(2)_A\times SU(2)_B\times U(1)_R}{U(1)_A\times U(1)_B}.
\end{equation}
Due to the fact that $\Wnk{2,4}$ is a non-symmetric coset \cite{Sakamoto2017} and can be realized as the double Wick rotation of the $\Tpq{1,1}$, integrability of \eqref{isometry} can not be ensured. Besides that, since a chaotic behaviour was found in $\Tpq{1,1}$ \cite{Basu2011} we could also expect it in $\Wnk{2,4}$ so that the $\sigma$ model on \eqref{isometry} would be non integrable. 
Notice that this coset has extra factors of $SO(2)_r\times U(1)_R$ in the numerator and two $U(1)$'s in the denominator in order to take into account all isometries of the space \cite{Crichigno2014,Rado2021a}

The Lie algebra that generates this space can then be written as
\begin{equation}
\label{algB}
\alggb:=\sl(2,\RR)^2\oplus\so(2)\oplus\su(2)^2\oplus\u(1)=\overbrace{\u(1)^4}^{\algi{0}}\oplus\overbrace{\left(\frac{\sl(2,\RR)^2\oplus\so(2)\oplus\su(2)^2\oplus\u(1)}{\u(1)^4}\right)}^{\algi{2}}.
\end{equation}
Taking a superalgebra that would include $\alggb$, its elements can be written as a supermatrix as
\begin{equation}
M_{(5|5)\times(5|5)}=\left(\begin{array}{c:cc|cc:c}
\u(1)_r & & & & & \\ 
\hdashline
 & \su(2)_A & & & Q & \\ 
 & & \su(2)_B & Q & & \\ 
\hline
 & & \bar{Q} & \sl(2,\RR) & & \\ 
 & \bar{Q} & & & \sl(2,\RR) & \\ 
\hdashline 
  & & & & & \u(1)_R  
\end{array}\right),
\end{equation}
where the dashed lines divide the supermatrix into blocks corresponding to the subspaces $\Wnk{2,4}$ and $\Tpq{1,1}$. $Q$ and $\bar{Q}$ belong to $\alggf=\algi{1}\oplus\algi{3}$ and correspond to the generators of the fermionic sector which will not be considered here.

We will use the same basis as in \cite{Sakamoto2017} to parametrize $\sl(2,\RR)^2\oplus\so(2)$ 
\begin{equation}
\label{gensW24}
\EL{\mu}{1}=\left(\begin{array}{c|cc:c}
0_{5\times 5} & & & \\ 
\hline
 & L_\mu & & \\ 
 & & 0_{2\times 2} & \\ 
\hdashline
 & & & 0_{1\times 1} 
\end{array}\right),
\EL{\mu}{2}=\left(\begin{array}{c|cc:c}
0_{5\times 5} & & & \\ 
\hline
 & 0_{2\times 2} & & \\ 
 & & L_\mu & \\ 
\hdashline
 & & & 0_{1\times 1} 
\end{array}\right),
\EN=-\frac{i}{2}\left(\begin{array}{c:c|c}
1\,& & \\ 
\hdashline 
 & 0_{4\times 4} &\\ 
\hline
 & & 0_{5\times 5} 
\end{array}\right),
\end{equation}
where $\mu=0,1,2,$ while for $\su(2)^2\oplus\u(1)$ we take 
\begin{equation}
\label{gensT11}
\EF{i}{1}=\left(\begin{array}{c:cc|c}
0_{1\times 1} & & & \\ 
\hdashline
 & F_i & & \\ 
 & & 0_{2\times 2} & \\ 
\hline
 & & & 0_{5\times 5} 
\end{array}\right),
\EF{i}{2}=\left(\begin{array}{c:cc|c}
0_{1\times 1} & & & \\ 
\hdashline
 & 0_{2\times 2} & & \\ 
 & & F_{i} & \\ 
\hline
 & & & 0_{5\times 5} 
\end{array}\right),
\EM=-\frac{i}{2}\left(\begin{array}{c|c:c}
0_{5\times 5} & & \\ 
\hline
 & 0_{4\times 4} &\\ 
\hdashline 
 & & \,1 
\end{array}\right)
\end{equation}
where $i=1,2,3$. Here, 
\begin{equation}
\label{Paulimats}
\begin{gathered}
L_0=\frac{i}{2}\es{3},\quad L_1=-\frac{1}{2}\es{2},\quad L_2=\frac{1}{2}\es{1},\\
F_1=-\frac{i}{2}\es{1},\quad F_2=-\frac{i}{2}\es{2},\quad F_3=-\frac{i}{2}\es{3},
\end{gathered}
\end{equation}
where $\es{i}$, $i=1,2,3$, are the standard Pauli matrices. The nonvanishing commutation rules are then 
\begin{equation}
\begin{gathered}
\comm{\EL{1}{m}}{\ELpm{n}}=\pm\dmn{mn}\ELpm{m},\quad\comm{\EL{+}{m}}{\EL{-}{n}}=2\dmn{mn}\EL{1}{m},\\
\comm{\EF{i}{m}}{\EF{j}{n}}=\dmn{mn}\eabc{ijk}\EF{k}{m},
\end{gathered}
\end{equation}
where $\ELpm{m}=\EL{2}{m}\pm\EL{0}{m}$. Notice that these commutation relations are those of two $\sl(2,\RR)$ and two $\su(2)$ for $m,n=1,2$. We also have 
\begin{equation}
\begin{gathered}
\Str{\EL{\mu}{m}\EL{\nu}{n}}=-\frac{1}{2}\dmn{mn}\eta_{\mu\nu},\quad\Str{\EF{i}{m}\EF{j}{n}}=-\frac{1}{2}\dmn{mn}\dij{ij},\\
\Str{\EN\EN}=-\Str{\EM\EM}=-\frac{1}{4},
\end{gathered}
\end{equation}
where $\eta_{\mu\nu}=\diag\left(-1,+1,+1\right)$. To parametrize \eqref{isometry} we follow \cite{Sakamoto2017} and define the basis
\begin{equation}
\label{gnum}
\frac{\sl(2,\RR)^2\oplus\so(2)\oplus\su(2)^2\oplus\u(1)}{\u(1)^4}=\text{span}\left\{\EL{1}{m},\EL{2}{m},i\sqrt{\frac{2}{3}}\EW,\EF{1}{m},\EF{2}{m},\sqrt{\frac{2}{3}}\EH\right\}=\left\{\EK{a}\right\},
\end{equation}
with
\begin{equation}
\Str{\EK{a}\EK{b}}=-\frac{1}{2}\dij{ab},
\end{equation}
where $m=1,2$ and $a=0,\dots,9,$ and 
\begin{equation}
\EW=\EL{0}{1}-\EL{0}{2}+\EN,\quad\EH=\EF{3}{1}-\EF{3}{2}+\EM.
\end{equation}
The four generators of $\u(1)^4$ are
\begin{equation}
\begin{gathered}
\label{gden}
\ET{1}=\EL{0}{1}+\EL{0}{2},\quad\ET{2}=\EL{0}{1}-\EL{0}{2}+4\EN,\\
\ET{3}=\EF{3}{1}+\EF{3}{2},\quad\ET{4}=\EF{3}{1}-\EF{3}{2}+4\EM.
\end{gathered}
\end{equation}

The undeformed metric of $\Wnk{2,4}\times\Tpq{1,1}$ can be obtained by choosing the following coset representatives for each subspace
\begin{equation}
\label{undefW24T11}
\begin{gathered}
g_{\Wnk{2,4}}=\exp\left(\vphi_1\EL{0}{1}+\vphi_2\EL{0}{2}+2\xi\EN\right)\exp\left((y_1-i\pi)\EL{1}{1}+y_2\EL{1}{2}\right),\\
g_{\Tpq{1,1}}=\exp\left(\phi_1\EF{3}{1}+\phi_2\EF{3}{2}+2\psi\EM\right)\exp\left((\th{1}+\pi)\EF{2}{1}+\th{2}\EF{2}{2}\right),
\end{gathered}
\end{equation}
which will lead us to a suitable form for the background. The Maurer-Cartan one form $A=g^{-1}dg$, where $g=g_{\Wnk{2,4}}g_{\Tpq{1,1}}$, allows us to get the metric from 
\begin{equation}
ds^2=-\frac{1}{3}\Str{A\Pb{A}}=ds^2_{\Wnk{2,4}}+ds^2_{\Tpq{1,1}}.
\end{equation}
To do that we take the projector $P_2$ on the coset to be
\begin{equation}
\label{P2}
\Pb{X}=\sum_{a=0}^9\frac{\Str{\EK{a}X}}{\Str{\EK{a}\EK{a}}}\EK{a}=X-\sum_{\a=1}^4\frac{\Str{\ET{\a}X}}{\Str{\ET{\a}\ET{\a}}}\ET{\a},
\end{equation}
where $\EK{a}$ and $\ET{\a}$ were defined in \eqref{gnum} and \eqref{gden}, respectively. We then obtain the following expressions for the undeformed metric on $\Wnk{2,4}\times\Tpq{1,1}$
\begin{equation}
\label{metricW24}
ds^2_{\Wnk{2,4}}=\frac{1}{6}\left(\dys{1}+\sinh^2\ys{1}\dvfs{1}\right)+\frac{1}{6}\left(\dys{2}+\sinh^2\ys{2}\dvfs{2}\right)-\frac{1}{9}\left(\cosh y_1\dvf{1}+\cosh y_2\dvf{2}+d\xi\right)^2,
\end{equation}
where $y_1,y_2\in[0,\infty)$, $\vf{1},\vf{2}\in[0,2\pi)$, and $\chi\in[0,4\pi)$, and also
\begin{equation}
\label{metricT11}
ds^2_{\Tpq{1,1}}=\frac{1}{6}\left(\dts{1}+\sin^2\th{1}\dfs{1}\right)+\frac{1}{6}\left(\dts{2}+\sin^2\th{2}\dfs{2}\right)+\frac{1}{9}\left(\cos\th{1}\df{1}+\cos\th{2}\df{2}+\dps\right)^2,
\end{equation}
where $\th{1},\th{2}\in\left[0,\pi\right]$, $\ph{1},\ph{2}\in[0,2\pi)$ and $\psi\in[0,4\pi)$. Notice the similarity between the subspaces. This happens because $\Wnk{2,4}$ can be seen as a $U(1)$-fibration over the product of two Euclidean $AdS_2$ spaces $EAdS_2\times EAdS_2$, while $\Tpq{1,1}$ can be understood as an $U(1)$-fiber over $\Sph{2}\times\Sph{2}$. 

\section{Bosonic \texorpdfstring{$\eta$}{eta}-deformed \texorpdfstring{$\Wnk{2,4}\times\Tpq{1,1}$}{W24xT11}\label{news}}
By switching off the fermionic degrees of freedom the deformed action \eqref{Sdef} reduces to
\begin{equation}
\label{YBdef}
S=-\frac{1}{3}
\left(1+\chi^2\right)
\int \dds\left(\g^{\a\b}-\varepsilon^{\a\b}\right)\Str{A_\a,\Pb{J_\b}},\quad\k=1.
\end{equation}
The action of $P_2$ on $A$, $\Rg{\EK{m}}$ and $J$ are
\begin{equation}
\label{P2AJ}
\Pb{A}=\Em{m}\EK{m},\quad\Pb{\Rg{\EK{m}}}=\Lmn{m}{n}\EK{n},\quad\Pb{J}=\jm{m}\EK{m},
\end{equation}
where $\EK{m}$ are the generators of $\algi{2}$ and the coefficients $\jm{m}$ can be obtained from
\begin{equation}
\label{jms}
\jm{m}=\EK{n}\Cmn{n}{m}.
\end{equation}
The coefficients $\Cmn{n}{m}$ and $\Lmn{m}{n}$ can be written in matrix form as
\begin{equation}
\label{Cmatrixdef}
\mathbf{C}=\left(\mathbf{I}-\chi\mathbf{\Lambda}\right)^{-1},\quad\chi=\frac{2\eta}{1-\eta^2}.
\end{equation}
Then, usign \eqref{YBdef} we can read off the metric and the B-field as 
\begin{equation}\label{Gmn}
ds^2=\Str{A\,\Pb{J}}=\jm{m}\Str{A\EK{m}}=\Em{m}\Cmn{m}{n}\Str{A\EK{n}},
\end{equation} 
\begin{equation}\label{Bmn}
B=\Str{A\wedge\Pb{J}}=-\jm{m}\wedge\Str{A\EK{m}}=\Em{m}\Cmn{m}{n}\wedge\Str{A\EK{n}}.
\end{equation}
Thus, in order to compute the deformed background with Lagrangian \eqref{YBdef} we need first to compute the the non-zero components $\Lmn{m}{n}$ in \eqref{P2AJ} and then follow the same procedure as in \cite{Rado2020,Rado2021a}. We find that
\begin{equation}
\begin{gathered}
\Lmn{1}{3}=-\Lmn{3}{1}=-\cosh y_1=q_1,\\
\Lmn{1}{5}=-\Lmn{5}{1}=-i\sqrt{\frac{2}{3}}\sinh y_1=q_2,\\
\Lmn{2}{4}=-\Lmn{4}{2}=\cosh y_2=q_3,\\
\Lmn{2}{5}=-\Lmn{5}{2}=-i\sqrt{\frac{2}{3}}\sinh y_2=q_4,\\
\Lmn{6}{8}=-\Lmn{8}{6}=-\cos\th{1}=q_5,\\
\Lmn{7}{9}=-\Lmn{9}{7}=\cos\th{2}=q_6,\\
\Lmn{8}{10}=-\Lmn{10}{8}=\sqrt{\frac{2}{3}}\sin\th{1}=q_7,\\
\Lmn{9}{10}=-\Lmn{10}{9}=\sqrt{\frac{2}{3}}\sin\th{2}=q_8.
\end{gathered}
\end{equation}
This allows to write the $\eta$-deformed background, where the $\Wnk{2,4}$ deformed is
\begin{equation}
\begin{gathered}
\label{metricW24def}
\frac{ds^2_{\Wnk{2,4}}}{\left(1+\chi^2\right)}=\Fd{1}\bigg(
\frac{1}{6}\big(\G{\qs{3}+\qs{4}}{0}\dys{1}+\G{\qs{1}+\qs{2}}{0}\dys{2}\\
+\G{\qs{2}+\qs{3}+\qs{4}}{\qs{2}\qs{3}}\sinh^2\!y_1\dvfs{1}+\G{\qs{1}+\qs{2}+\qs{4}}{\qs{1}\qs{4}}\sinh^2\!y_2\dvfs{2}\big)\\
-\frac{1}{9}\G{\qs{1}}{0}\G{\qs{3}}{0}\left(\cosh y_1\dvf{1}+\cosh y_2\dvf{2}+\dch\right)^2
\bigg),
\end{gathered}
\end{equation}
while the deformed metric of $\Tpq{1,1}$ is 
\begin{equation}
\begin{gathered}
\label{metricT11def}
\frac{ds^2_{\Tpq{1,1}}}{\left(1+\chi^2\right)}=\Fd{2}\bigg(
\frac{1}{6}\big(\G{\qs{6}+\qs{8}}{0}\dts{1}+\G{\qs{5}+\qs{7}}{0}\dts{2}\\
+\G{\qs{6}+\qs{7}+\qs{8}}{\qs{6}\qs{7}}\sin^2\!\th{1}\dfs{1}+\G{\qs{5}+\qs{7}+\qs{8}}{\qs{5}\qs{8}}\sin^2\!\th{2}\dfs{2}\big)\\
+\frac{1}{9}\G{\qs{5}}{0}\G{\qs{6}}{0}\left(\cos\th{1}\df{1}+\cos\th{2}\df{2}+\dps\right)^2
\bigg),
\end{gathered}
\end{equation}
where 
\begin{equation}
\begin{gathered}
\label{fdef}
\Fdi{1}=\G{\qs{1}+\qs{2}+\qs{3}+\qs{4}}{\qs{5}\qs{6}+\qs{5}\qs{8}+\qs{6}\qs{7}},\\
\Fdi{2}=\G{\qs{5}+\qs{6}+\qs{7}+\qs{8}}{\qs{5}\qs{6}+\qs{5}\qs{8}+\qs{6}\qs{7}},
\end{gathered}
\end{equation}
and
\begin{equation}
\label{Gdef}
\G{r}{s}=1+r\chi^2+s\chi^4,
\end{equation}
with $\Fdi{1}=\Fdi{2}=1$ when $\chi=0$. Notice that
when we set $\chi=0$ the deformed metrics \eqref{metricW24def} and \eqref{metricT11def} reduce to the undeformed ones. 

The deformed NSNS $B$-field is then given by 
\begin{equation}
\label{BfieldW24}
\begin{gathered}
\frac{B_{\Wnk{2,4}}}{\left(1+\chi^2\right)}=
+\frac{i\chi\Fd{1}}{9}\bigg(
\left(\sqrt{6}q_2\G{\qs{3}}{0}\cosh y_1-3iq_1\G{\qs{3}+\qs{4}}{0}\sinh y_1\right)dy_1\wedge\dvf{1}\\
+q_2\left(\sqrt{6}\G{\qs{3}}{0}\cosh y_2-3iq_3q_4\chi^2\sinh y_2\right)dy_1\wedge\dvf{2}\\
+q_4\left(\sqrt{6}\G{\qs{1}}{0}\cosh dy_1+3iq_1q_2\chi^2\sinh y_1\right)dy_2\wedge\dvf{1}\\
+\left(\sqrt{6}q_4\G{\qs{1}}{0}\cosh y_2+3iq_3\G{\qs{1}+\qs{2}}{0}\sinh y_2\right)dy_2\wedge\dvf{2}\\
+\sqrt{6}q_2\G{\qs{3}}{0}dy_1\wedge\dch+\sqrt{6}q_4\G{\qs{1}}{0}dy_2\wedge\dch\bigg),
\end{gathered}
\end{equation}
\begin{equation}
\label{BfieldT11}
\begin{gathered}
\frac{B_{\Tpq{1,1}}}{\left(1+\chi^2\right)}=-\frac{\chi\Fd{2}}{9}\bigg(
\left(\sqrt{6}q_7\G{\qs{6}}{0}\cos\th{1}+3q_5\G{\qs{6}+\qs{8}}{0}\sin\th{1}\right)\dth{1}\wedge\df{1}\\
+q_7\left(\sqrt{6}\G{\qs{6}}{0}\cos\th{2}+3q_6q_8\chi^2\sin\th{2}\right)\dth{1}\wedge\df{2}\\
+q_8\left(\sqrt{6}\G{\qs{5}}{0}\cos\th{1}-3q_5q_7\chi^2\sin\th{1}\right)\dth{2}\wedge\df{1}\\
+\left(\sqrt{6}q_8\G{\qs{5}}{0}\cos\th{2}-3q_6\G{\qs{5}+\qs{7}}{0}\sin\th{2}\right)\dth{2}\wedge\df{2}\\
+\sqrt{6}q_7\G{\qs{6}}{0}\dth{1}\wedge\dps+\sqrt{6}q_8\G{\qs{5}}{0}\dth{2}\wedge\dps\bigg),
\end{gathered}
\end{equation}
which vanishes when we set $\chi=0$.

{\section{Coset Construction of the Bosonic \texorpdfstring{$\AdS{5}\times\Tpq{1,1}$}{AdS5xT11} Background \label{cosetconstruction2}}
We can now consider $\AdS{5}\times\Tpq{1,1}$ as a second example of non-integrable background \cite{Basu2011}. It was used in \cite{Crichigno2014} and recently in \cite{Rado2021a} to build Yang-Baxter deformations based on solutions of the CYBE.  This background corresponds to the near horizon limit of $N$ $D3$-branes on the singularity of $\mathcal{M}_{1,4}\times Y_6$ where $\mathcal{M}_{1,4}$ is the four-dimensional Minkowski space and $Y_6$ a Ricci flat Calabi-Yau cone $C(X_5)$ with base $X_5$ \cite{Klebanov1998}. In the limit the geometry becomes $\AdS{5}\times X_5$, where $X_5$ is a compact Sasaki-Einstein manifold which, in particular, can be taken to be $\Tpq{1,1}$ conifold.

In this case the isometry group is the extended coset \cite{Crichigno2014,Rado2021a}
\begin{equation}
\label{isometry2}
\AdS{5}\times\Tpq{1,1}\equiv\frac{SO(2,4)}{SO(1,4)}\times\frac{SU(2)_A\times SU(2)_B\times U(1)_R}{U(1)_A\times U(1)_B},
\end{equation}
which leads to the following extended coset algebra:
\begin{equation}
\label{cosetT11}
\so(2,4)\oplus\su(2)\oplus\su(2)\oplus\u(1)=\overbrace{\left(\so(1,4)\oplus\u(1)\oplus\u(1)\right)}^{\algi{0}}\oplus\overbrace{\left(\frac{\so(2,4)\oplus\su(2)\oplus\su(2)\oplus\u(1)_R}{\so(1,4)\oplus\u(1)\oplus\u(1)}\right)}^{\algi{2}=\algg/\algi{0}}.
\end{equation}
We can consider a slightly different supermatrix structure
\begin{equation}
\label{supermatrix2}
M_{(4|5)\times(4|5)}=\left(\begin{array}{cc|c:c}
\su(2)_A & & & \\ 
 & \su(2)_B & & \\ 
\hline
 & & \u(1)_R & \\ 
\hdashline 
 & & & \so(2,4) 
\end{array}\right),
\end{equation}
where the dashed lines split the algebras corresponding to the subspaces $\AdS{5}$ and $\Tpq{1,1}$, while the solid lines split the $M_{4\times 4}$ and $M_{5\times 5}$ bosonic blocks.

The basis of $\su(2)^2\oplus\u(1)$ generators which parametrize $\Tpq{1,1}$ is the following
\begin{equation}
\label{gensT11_2}
\EF{i}{1}=\left(\begin{array}{cc|c}
F_i & & \\ 
& 0_{2\times 2} & \\ 
\hline
& & 0_{5\times 5} 
\end{array}\right),
\EF{i}{2}=\left(\begin{array}{cc|c}
0_{2\times 2} & & \\ 
& F_{i} & \\ 
\hline
& & 0_{5\times 5} 
\end{array}\right),
\EM=-\frac{i}{2}\left(\begin{array}{c|c:c}
0_{4\times 4} & & \\ 
\hline
 & \,1 &\\ 
\hdashline 
 & & 0_{4\times 4}
\end{array}\right)
\end{equation}
where $i=1,2,3$ and $F_i$ are given in \eqref{Paulimats}. On the other hand, we take the basis of $\so(2,4)$ in terms of $\g$-matrices in a similar way as done in \cite{Rado2021a},
\begin{equation}
\begin{gathered}
{\bf\Gamma}_\mu=\left(\begin{array}{c|c:c}
0_{4\times 4} & & \\ 
\hline
 & 0 &\\ 
\hdashline 
 & & \g_\mu
\end{array}\right),
{\bf\Gamma}_5=\left(\begin{array}{c|c:c}
0_{4\times 4} & & \\ 
\hline
 & 0 &\\ 
\hdashline 
 & & \g_5
\end{array}\right),\\
\EMa{\mu\nu}=\left(\begin{array}{c|c:c}
0_{4\times 4} & & \\ 
\hline
 & 0 &\\ 
\hdashline 
 & & m_{\mu\nu}
\end{array}\right),
\EMa{\mu 5}=\left(\begin{array}{c|c:c}
0_{4\times 4} & & \\ 
\hline
 & 0 &\\ 
\hdashline 
 & & m_{\mu 5}
\end{array}\right),
\end{gathered}
\end{equation}
where $\g_\mu$, $\g_5$, $m_{\mu\nu}=1/4\comm{\g_\mu}{\g_\nu}$ and $m_{\mu5}=1/4\comm{\g_\mu}{\g_5}$ are the fifteen $4\times 4$ matrices for the generators of isometries of $\AdS{5}$. 

To parametrize \eqref{isometry2} we set the basis
\begin{equation}
\frac{\su(2)^2\oplus\u(1)\oplus\so(2,4)}{\u(1)^2\oplus\so(1,4)}=\left\{\Km{m}\right\},\quad m=0,\dots,9,
\end{equation}
where 
\begin{equation}
\begin{gathered}
\Km{0}=\frac{1}{2}{\bf\Gamma}_0,\quad\Km{1}=\frac{1}{2}{\bf\Gamma}_1,\quad\Km{2}=\frac{1}{2}{\bf\Gamma}_2,\quad\Km{3}=\frac{1}{2}{\bf\Gamma}_3,\quad\Km{4}=\frac{1}{2}{\bf\Gamma}_5,\\
\Km{5}=\EF{1}{1},\quad\Km{6}=\EF{1}{2},\quad\Km{7}=\EF{2}{1},\quad\Km{8}=\EF{2}{2}\quad\Km{9}=\sqrt{\frac{2}{3}}\EH.
\end{gathered}
\end{equation}
The undeformed metric $\AdS{5}$ can be obtained by choosing the following coset representative
\begin{equation}
g_{\AdS{5}}=\exp\frac{i}{2}\left(\psi_1\EHi{1}+\psi_2\EHi{2}+t\EHi{3}\right)\exp\left(-\zeta\EMa{1,3}\right)\exp\frac{1}{2}\left(-\sinh^{-1}\!\!\rho{\bf\Gamma}_1\right),
\end{equation}
while the coset representative for $g_{\Tpq{1,1}}$ is given in \eqref{undefW24T11}. Using the projector \eqref{P2} we obtain the following expression for the undeformed $\AdS{5}$ metric
\begin{equation}
\label{metricAdS5}
ds^2_{\AdS{5}}=-(1+\rho^2)dt^2+\frac{d\rho^2}{1+\rho^2}+\rho^2(d\zeta^2+\cos^2\zeta\,d\psi_1^2+\sin^2\zeta\,d\psi_2^2),
\end{equation}
which agrees with that in \cite{Arutyunov2014,Arutyunov2015}. Finally, the undeformed $\Tpq{1,1}$ is written as in \eqref{metricT11}.
}

{\section{Bosonic \texorpdfstring{$\eta$}{eta}-deformed \texorpdfstring{$\AdS{5}\times\Tpq{1,1}$}{AdS5xT11}\label{news2}}
In this section we obtain the $\eta$-deformation of $\AdS{5}\times\Tpq{1,1}$ based on the $R$ operator \eqref{RMij}. As in \autoref{news}, we switch off the fermionic degrees of freedom to work only with the bosonic sector. Firstly, we need to rewrite the components of the ${\bf\Lambda}$ matrix from \eqref{P2AJ}, which in this case takes the form 
\begin{equation}
\label{Lij2}
\begin{gathered}
\Lmn{2}{5}=-\Lmn{5}{2}=i\rho=q_1,\\
\Lmn{3}{4}=-\Lmn{4}{3}=-\rho^2\sin\zeta=q_2,\\
\Lmn{6}{8}=-\Lmn{8}{6}=-\cos\theta_1=q_3,\\
\Lmn{7}{9}=-\Lmn{9}{7}=+\cos\theta_2=q_4,\\
\Lmn{8}{10}=-\Lmn{10}{8}=\sqrt{\frac{2}{3}}\sin\theta_1=q_5,\\
\Lmn{9}{10}=-\Lmn{10}{9}=\sqrt{\frac{2}{3}}\sin\theta_2=q_6,
\end{gathered}
\end{equation}
which allow us to construct the  matrix ${\bf C}$ in \eqref{Cmatrixdef}. 

This also allow us to write the $\eta$-deformed background with the $\AdS{5}$ deformed space being 
\begin{equation}
\label{metricAdS5def}
\frac{ds_{\AdS{5}}^2}{(1+\chi^2)}=-\frac{(1+\rho^2)dt^2}{1+\chi^2q_1^2}+\frac{d\rho^2}{(1+\rho^2)(1+\chi^2q_1^2)}+\rho^2\left(\frac{d\zeta^2+\cos^2\zeta\,d\psi_1^2}{1+\chi^2q_2^2}+\sin^2\zeta\,d\psi_2^2\right),
\end{equation}
agreeing with the $\eta$-deformed $\AdS{5}$ space obtained in \cite{Arutyunov2014,Arutyunov2015}. The $\eta$-deformed $\Tpq{1,1}$ metric is
\begin{equation}
\label{metricT11def2}
\begin{gathered}
\frac{ds^2_{\Tpq{1,1}}}{\left(1+\chi^2\right)}=\Fd{3}\bigg(
\frac{1}{6}\big(\G{\qs{4}+\qs{6}}{0}\dts{1}+\G{\qs{3}+\qs{5}}{0}\dts{2}\\
+\G{\qs{4}+\qs{5}+\qs{6}}{\qs{4}\qs{5}}\sin^2\!\th{1}\dfs{1}+\G{\qs{3}+\qs{5}+\qs{6}}{\qs{3}\qs{6}}\sin^2\!\th{2}\dfs{2}\big)\\
+\frac{1}{9}\G{\qs{3}}{0}\G{\qs{4}}{0}\left(\cos\th{1}\df{1}+\cos\th{2}\df{2}+\dps\right)^2
\bigg),
\end{gathered}
\end{equation}
where 
\begin{equation}
\Fdi{3}=\G{\qs{3}+\qs{4}+\qs{5}+\qs{6}}{\qs{3}\qs{4}+\qs{3}\qs{6}+\qs{4}\qs{5}}.
\end{equation}
where we have used the definition \eqref{Gdef}. Notice that this latter deformation is the same as in \eqref{metricT11def}. Both \eqref{metricAdS5def} and \eqref{metricT11def2} reduce to the undeformed spaces when $\chi=0$.

Due to the $\eta$-deformation a $B$-field is also present in each sector
\begin{equation}
\label{BfieldAdS5}
\begin{gathered}
\frac{B_{\AdS{5}}}{(1+\chi^2)}=\frac{2i\chi q_1^2}{3(1+\chi^2q_1)}dt\wedge d\rho-\frac{2\chi q_2\rho\cos\zeta}{3(1+\chi^2q_2^2)}d\zeta\wedge d\psi_1,
\end{gathered}
\end{equation}
which is proportional to the $B$-field with $\AdS{5}$ coordinates in \cite{Arutyunov2014,Arutyunov2015}, and 
\begin{equation}
\label{BfieldT112}
\begin{gathered}
\frac{B_{\Tpq{1,1}}}{\left(1+\chi^2\right)}=-\frac{\chi\Fd{3}}{9}\bigg(
\left(\sqrt{6}q_5\G{\qs{4}}{0}\cos\th{1}+3q_3\G{\qs{4}+\qs{6}}{0}\sin\th{1}\right)\dth{1}\wedge\df{1}\\
+q_5\left(\sqrt{6}\G{\qs{4}}{0}\cos\th{2}+3q_4q_6\chi^2\sin\th{2}\right)\dth{1}\wedge\df{2}\\
+q_6\left(\sqrt{6}\G{\qs{5}}{0}\cos\th{1}-3q_3q_5\chi^2\sin\th{1}\right)\dth{2}\wedge\df{1}\\
+\left(\sqrt{6}q_6\G{\qs{3}}{0}\cos\th{2}-3q_4\G{\qs{3}+\qs{5}}{0}\sin\th{2}\right)\dth{2}\wedge\df{2}
\\
+\sqrt{6}q_5\G{\qs{4}}{0}\dth{1}\wedge\dps+\sqrt{6}q_6\G{\qs{3}}{0}\dth{2}\wedge\dps\bigg).
\end{gathered}
\end{equation}
The $B$-field in \eqref{BfieldT112} has the same form as the one found previously \eqref{BfieldT11}. As expected, these $B$-fields vanish for $\chi=0$.
}
\section{Conclusions \label{conclusions}}
In this paper we have obtained new bosonic $\eta$-deformed $\Wnk{2,4}\times\Tpq{1,1}$ and $\AdS{5}\times\Tpq{1,1}$ backgrounds based on a solution of the mCYBE  providing new integrable backgrounds besides the one found in \cite{Sakamoto2017} for $\Wnk{2,4}\times\Tpq{1,1}$.


Recently new integrable models on $\Tpq{1,1}$ manifolds with non-vanishing B-fields were found \cite{Arutyunov2020}. It was then shown that such B-fields are the only possible ones which avoid a chaotic behaviour of $\Tpq{1,1}$ \cite{Ishii2021}. Since $\Wnk{2,4}$ is closely related to  $\Tpq{1,1}$ by two Wick rotations we would expect that it gives rise to a  $B$-field as in  \cite{Arutyunov2020} and  it would also be very interesting to look for integrable deformations of $\Wnk{2,4}\times\Tpq{1,1}$ along the lines of \cite{Arutyunov2020}. Besides that, it would also be interesting to analyse whether our deformed $\Wnk{2,4}\times\Tpq{1,1}$ and $\AdS{5}\times\Tpq{1,1}$ backgrounds present chaotic behaviour.


\acknowledgments
The work of V.O. Rivelles was supported by FAPESP grant  2019/21281-4.

\bibliographystyle{JHEP}
\bibliography{Biblioteca}

\providecommand{\href}[2]{#2}\begingroup\raggedright\begin{thebibliography}{10}

\bibitem{Maillet1985}
J.-M.~Maillet, \emph{Kac-$\mathrm{M}$oody algebra and extended yang-baxter
  relations in the $\mathrm{O(N)}$ non-linear $\sigma$-model},
  \href{https://doi.org/https://doi.org/10.1016/0370-2693(85)91075-5}{\emph{Physics
  Letters B} {\bfseries 162} (1985) 137}.

\bibitem{Maillet1986}
J.-M.~Maillet, \emph{New integrable canonical structures in two-dimensional
  models},
  \href{https://doi.org/http://dx.doi.org/10.1016/0550-3213(86)90365-2}{\emph{Nuclear
  Physics B} {\bfseries 269} (1986) 54}.

\bibitem{Maillet1986a}
J.-M.~Maillet, \emph{{Hamiltonian structures for integrable classical theories
  from graded Kac-Moody algebras}},
  \href{https://doi.org/https://doi.org/10.1016/0370-2693(86)91289-X}{\emph{Physics
  Letters B} {\bfseries 167} (1986) 401}.

\bibitem{Bena2004}
I.~Bena, J.~Polchinski and R.~Roiban, \emph{{Hidden symmetries of the AdS(5) x
  S**5 superstring}},
  \href{https://doi.org/10.1103/PhysRevD.69.046002}{\emph{Phys. Rev. D}
  {\bfseries 69} (2004) 046002}
  [\href{https://arxiv.org/abs/hep-th/0305116}{{\ttfamily hep-th/0305116}}].

\bibitem{Magro2009}
M.~Magro, \emph{{The Classical Exchange Algebra of $AdS_5\times S^5$}},
  \href{https://doi.org/10.1088/1126-6708/2009/01/021}{\emph{JHEP} {\bfseries
  01} (2009) 021} [\href{https://arxiv.org/abs/0810.4136}{{\ttfamily
  0810.4136}}].

\bibitem{Cherednik1981}
I.V.~Cherednik, \emph{{Relativistically Invariant Quasiclassical Limits of
  Integrable Two-dimensional Quantum Models}},
  \href{https://doi.org/10.1007/BF01086395}{\emph{Theor. Math. Phys.}
  {\bfseries 47} (1981) 422}.

\bibitem{Klimcik2009}
C.~Klimcik, \emph{{On integrability of the Yang-Baxter sigma-model}},
  \href{https://doi.org/10.1063/1.3116242}{\emph{J. Math. Phys.} {\bfseries 50}
  (2009) 043508} [\href{https://arxiv.org/abs/0802.3518}{{\ttfamily
  0802.3518}}].

\bibitem{Delduc2013}
F.~Delduc, M.~Magro and B.~Vicedo, \emph{{On classical $q$-deformations of
  integrable sigma-models}},
  \href{https://doi.org/10.1007/JHEP11(2013)192}{\emph{JHEP} {\bfseries 11}
  (2013) 192} [\href{https://arxiv.org/abs/1308.3581}{{\ttfamily 1308.3581}}].

\bibitem{Delduc2014}
F.~Delduc, M.~Magro and B.~Vicedo, \emph{{An integrable deformation of the
  $AdS_5\times S^{5}$ superstring action}},
  \href{https://doi.org/10.1103/PhysRevLett.112.051601}{\emph{Phys. Rev. Lett.}
  {\bfseries 112} (2014) 051601}
  [\href{https://arxiv.org/abs/1309.5850}{{\ttfamily 1309.5850}}].

\bibitem{Delduc2014a}
F.~Delduc, M.~Magro and B.~Vicedo, \emph{{Derivation of the action and
  symmetries of the $q$-deformed $AdS_{5} \times S^{5}$ superstring}},
  \href{https://doi.org/10.1007/JHEP10(2014)132}{\emph{JHEP} {\bfseries 10}
  (2014) 132} [\href{https://arxiv.org/abs/1406.6286}{{\ttfamily 1406.6286}}].

\bibitem{Arutyunov2014}
G.~Arutyunov, R.~Borsato and S.~Frolov, \emph{{S-matrix for strings on
  $\eta$-deformed $AdS_5\times S^5$}},
  \href{https://doi.org/10.1007/JHEP04(2014)002}{\emph{JHEP} {\bfseries 04}
  (2014) 002} [\href{https://arxiv.org/abs/1312.3542}{{\ttfamily 1312.3542}}].

\bibitem{Arutyunov2015}
G.~Arutyunov, R.~Borsato and S.~Frolov, \emph{{Puzzles of $\eta$-deformed
  $AdS_5 \times S^5$}},
  \href{https://doi.org/10.1007/JHEP12(2015)049}{\emph{JHEP} {\bfseries 12}
  (2015) 049} [\href{https://arxiv.org/abs/1507.04239}{{\ttfamily
  1507.04239}}].

\bibitem{Rado2021}
L.~Rado, V.O.~Rivelles and R.~S\'anchez, \emph{{Bosonic $\eta$-deformed
  $AdS_4\times\mathbb{CP}^3$ Background}},
  \href{https://doi.org/10.1007/JHEP10(2021)115}{\emph{JHEP} {\bfseries 10}
  (2021) 115} [\href{https://arxiv.org/abs/2105.07545}{{\ttfamily
  2105.07545}}].

\bibitem{Matsumoto2014}
T.~Matsumoto and K.~Yoshida, \emph{{Lunin-Maldacena backgrounds from the
  classical Yang-Baxter equation - towards the gravity/CYBE correspondence}},
  \href{https://doi.org/10.1007/JHEP06(2014)135}{\emph{JHEP} {\bfseries 06}
  (2014) 135} [\href{https://arxiv.org/abs/1404.1838}{{\ttfamily 1404.1838}}].

\bibitem{Matsumoto2014a}
T.~Matsumoto and K.~Yoshida, \emph{{Integrability of classical strings dual for
  noncommutative gauge theories}},
  \href{https://doi.org/10.1007/JHEP06(2014)163}{\emph{JHEP} {\bfseries 06}
  (2014) 163} [\href{https://arxiv.org/abs/1404.3657}{{\ttfamily 1404.3657}}].

\bibitem{Matsumoto2015a}
T.~Matsumoto and K.~Yoshida, \emph{{Schrödinger geometries arising from
  Yang-Baxter deformations}},
  \href{https://doi.org/10.1007/JHEP04(2015)180}{\emph{JHEP} {\bfseries 04}
  (2015) 180} [\href{https://arxiv.org/abs/1502.00740}{{\ttfamily
  1502.00740}}].

\bibitem{Negron2018}
R.~Negrón and V.O.~Rivelles, \emph{{Yang-Baxter deformations of the
  $AdS_4\times\mathbb{CP}^3$ superstring sigma model}},
  \href{https://doi.org/10.1007/JHEP11(2018)043}{\emph{JHEP} {\bfseries 11}
  (2018) 043} [\href{https://arxiv.org/abs/1809.01174}{{\ttfamily
  1809.01174}}].

\bibitem{Rado2020}
L.~Rado, V.O.~Rivelles and R.~S\'anchez, \emph{{String backgrounds of the
  Yang-Baxter deformed $AdS_4\times\mathbb{CP}^3$ superstring}},
  \href{https://doi.org/10.1007/JHEP01(2021)056}{\emph{JHEP} {\bfseries 01}
  (2021) 056} [\href{https://arxiv.org/abs/2009.04397}{{\ttfamily
  2009.04397}}].

\bibitem{Crichigno2014}
P.M.~Crichigno, T.~Matsumoto and K.~Yoshida, \emph{{Deformations of $T^{1,1}$
  as Yang-Baxter sigma models}},
  \href{https://doi.org/10.1007/JHEP12(2014)085}{\emph{JHEP} {\bfseries 12}
  (2014) 085} [\href{https://arxiv.org/abs/1406.2249}{{\ttfamily 1406.2249}}].

\bibitem{Rado2021a}
L.~Rado, V.O.~Rivelles and R.~S\'anchez, \emph{{Yang-Baxter deformations of the
  $AdS_5$ x $T^{1,1}$ superstring and their backgrounds}},
  \href{https://doi.org/10.1007/JHEP02(2021)126}{\emph{JHEP} {\bfseries 02}
  (2021) 126} [\href{https://arxiv.org/abs/2010.14081}{{\ttfamily
  2010.14081}}].

\bibitem{Sakamoto2017}
J.-i.~Sakamoto and K.~Yoshida, \emph{{Yang-Baxter deformations of
  $W_{2,4}\times T^{1,1}$ and the associated T-dual models}},
  \href{https://doi.org/10.1016/j.nuclphysb.2017.06.017}{\emph{Nucl. Phys. B}
  {\bfseries 921} (2017) 805}
  [\href{https://arxiv.org/abs/1612.08615}{{\ttfamily 1612.08615}}].

\bibitem{BrittoPacumio1999}
R.~Britto-Pacumio, A.~Strominger and A.~Volovich, \emph{{Holography for coset
  spaces}}, \href{https://doi.org/10.1088/1126-6708/1999/11/013}{\emph{JHEP}
  {\bfseries 11} (1999) 013}
  [\href{https://arxiv.org/abs/hep-th/9905211}{{\ttfamily hep-th/9905211}}].

\bibitem{PandoZayas2000}
L.A.~Pando~Zayas and A.A.~Tseytlin, \emph{{Conformal sigma models for a class
  of T**(p,q) spaces}},
  \href{https://doi.org/10.1088/0264-9381/17/24/312}{\emph{Class. Quant. Grav.}
  {\bfseries 17} (2000) 5125}
  [\href{https://arxiv.org/abs/hep-th/0007086}{{\ttfamily hep-th/0007086}}].

\bibitem{Arutyunov2020}
G.~Arutyunov, C.~Bassi and S.~Lacroix, \emph{{New integrable coset sigma
  models}}, \href{https://doi.org/10.1007/JHEP03(2021)062}{\emph{JHEP}
  {\bfseries 03} (2021) 062}
  [\href{https://arxiv.org/abs/2010.05573}{{\ttfamily 2010.05573}}].

\bibitem{Basu2011}
P.~Basu and L.A.~Pando~Zayas, \emph{{Chaos rules out integrability of strings
  on AdS$_5 \times T^{1,1}$}},
  \href{https://doi.org/10.1016/j.physletb.2011.04.063}{\emph{Phys. Lett. B}
  {\bfseries 700} (2011) 243}
  [\href{https://arxiv.org/abs/1103.4107}{{\ttfamily 1103.4107}}].

\bibitem{Vicedo2015}
B.~Vicedo, \emph{{Deformed integrable $\sigma$-models, classical R-matrices and
  classical exchange algebra on Drinfeld doubles}},
  \href{https://doi.org/10.1088/1751-8113/48/35/355203}{\emph{J. Phys.}
  {\bfseries A48} (2015) 355203}
  [\href{https://arxiv.org/abs/1504.06303}{{\ttfamily 1504.06303}}].

\bibitem{James1976}
I.~James, C.U.~Press and N.~Hitchin, \emph{The Topology of Stiefel Manifolds},
  Cambridge books online, Cambridge University Press (1976).

\bibitem{Klebanov1998}
I.R.~Klebanov and E.~Witten, \emph{{Superconformal field theory on three-branes
  at a Calabi-Yau singularity}},
  \href{https://doi.org/10.1016/S0550-3213(98)00654-3}{\emph{Nucl. Phys. B}
  {\bfseries 536} (1998) 199}
  [\href{https://arxiv.org/abs/hep-th/9807080}{{\ttfamily hep-th/9807080}}].

\bibitem{Ishii2021}
T.~Ishii, S.~Kushiro and K.~Yoshida, \emph{{Chaotic string dynamics in deformed
  T$^{1,1}$}}, \href{https://doi.org/10.1007/JHEP05(2021)158}{\emph{JHEP}
  {\bfseries 05} (2021) 158}
  [\href{https://arxiv.org/abs/2103.12416}{{\ttfamily 2103.12416}}].

\end{thebibliography}\endgroup
\end{document}